# Free-space transfer of comb-rooted optical frequencies over an 18 km open-air link

Hyun Jay Kang[1], Jae-Won Yang[1], Byung Jae Chun[1,2,3], Heesuk Jang[1,4], Byung Soo Kim[1], Young-Jin Kim[2,*] and Seung-Woo Kim[1,*]

[1]Department of Mechanical Engineering, Korea Advanced Institute of Science and Technology (KAIST), Science Town, Daejeon, 34141, South Korea.

[2]School of Mechanical and Aerospace Engineering, Nanyang Technological University (NTU), 50 Nanyang Avenue, Singapore 639798, Singapore.

[3]Presently with Department of Quantum Optics Division, Korea Atomic Energy Research Institute (KAERI), Science Town, Daejeon, 34057, South Korea.

[4]Presently with Agency for Defense Development (ADD), Yuseong, Daejeon, 34186, South Korea.

*Corresponding authors: swk@kaist.ac.kr and yj.kim@ntu.edu.sg

**Phase-coherent transfer of optical frequencies over a long distance is required for diverse photonic applications, including optical clock signal dissemination and physical constants measurement. Several demonstrations have been made successfully over fiber networks, but not much work has been done yet through the open air where atmospheric turbulence prevails. Here, via an 18 km outdoor link, we transmit multiple optical carriers that are extracted directly from a near-infrared frequency comb over a 4.2 THz spectral range in stabilization to a high-finesse cavity with a 1.5 Hz linewidth. Proof-of-concept experiments show that the comb-rooted optical carriers are transferred in parallel with collective suppression of atmospheric phase noise to -80 dBc/Hz. In addition, microwaves can also be delivered by pairing two separate optical carriers bound with inter-comb-mode coherence, e.g. a 10 GHz signal with phase noise of -105 dBc/Hz at 1 Hz offset. Further, an add-on demonstration is made on multi-channel coherent optical communications with the potential of multi-Tbps data transmission in free space.**

Long-distance transfer of phase-coherent optical frequencies is needed for many photonic applications such as optical clock signal dissemination[1,2], fundamental physical constants measurements[3-5], astronomical long-baseline interferometers[6,7], and coherent optical communications[8]. Fiber networks prove to be a reliable optical link capable of delivering optical frequencies over several hundreds of kilometers while suppressing phase noise down to $10^{-18}$ at 1000 s integration[9,10]. Nonetheless, there are strong demands on free-space optical links where no fiber network is available for practical reasons. Particularly, in order to advance ground-to-satellite applications, open-air optical links are required to supplement present-day radio-frequency and microwave transmissions, as attempted for the time transfer by laser link (T2L2)[11], optical two-way time and frequency transfer (TWTFT)[12-14], and lunar laser communications demonstration (LLCD)[15]. In the air, however, light propagation is greatly affected by atmospheric turbulence stochastically in forms of intensity scintillation and phase fluctuation[16], which significantly limits the practical range of coherent optical frequency transfer when the Doppler shift is not actively compensated[17-19].

In the meanwhile, the frequency comb technology has made a remarkable progress to produce highly stable optical frequencies with direct stabilization to the microwave or optical atomic clocks[20,21]. It is now anticipated to play a role in the advancement of free-space optical links, as already evident for comb-based TWTFT[12-14] and massive parallel coherent optical communications[22,23]. In this context, here we demonstrate a comb-rooted scheme of transferring multiple optical carriers concurrently over an 18 km open-air optical link. The optical carriers are generated from a near-infrared frequency comb over a 35 nm wavelength range, or equivalently a 4.2 THz span, in stabilization to a high-finesse cavity with a narrow linewidth of 1.5 Hz at 1 s integration. The optical link is built long enough to simulate the turbulence effect encountered over the troposphere in ground-to-satellite transmissions[18,19]. The subsequent transfer of the comb-rooted optical carriers over the long open-air link is made to reveal that the atmospheric phase disturbance adheres to the weak-turbulence model even with large Rytov variance[16], thereby the Doppler shift can be suppressed actively in the presence of intensity scintillation. Moreover, the concurrent transfer of multiple comb-rooted optical carriers is intended for long-distance delivery of not just individual optical frequencies but also microwaves that are synthesized by paring different optical carriers bound with high inter-comb-mode coherence.

Figure 1 illustrates the overall transmission system configured in this study for free-space transfer of comb-rooted optical carriers over a long distance. The optical link is set up on the KAIST main campus between Site A to Site B, with a flat mirror of a 300 mm diameter being placed to fold the optical path halfway on the top of a nearby mountain. Site B is consequently situated in close proximity to Site A to facilitate the direct comparative evaluation of the optical carriers upon arrival in comparison to their original signals. At Site A, the optical carriers are generated directly from the frequency comb of an Er-doped fiber oscillator with stabilization to a high-finesse cavity. All the generated carriers are then delivered together from the source oscillator to Site A through a single-mode fiber delivery line of a 250 m length, with optical power being boosted to 1.0 W through an Er-doped fiber amplifier (EDFA). Open-air transmission is made through a refractive-type telescope having an 80 mm aperture diameter. The optical carriers arriving at Site B are collected through a reflective Cassegrain-type telescope of an aperture diameter of 250 mm. This bistatic transmitter-to-receiver telescopic configuration is pre-aligned to maximize the optical power transmission. In addition, a fine tracking control mechanism (Methods) is installed in the receiver telescope to reduce the intensity fluctuation due to beam wandering. As a result, on mild weather conditions, the optical power focused on the receiving fiber at Site B maintains an average of 2.7 µW (Fig. 1c-1). The power attenuation of about 55 dB is attributable mainly to the absorption and scattering caused by atmospheric particulates[24], including a 10 dB power loss during light collimation on to the single-mode receiving fiber from free space.

The optical power received at Site B suffers a dynamic fluctuation that is observed dominant for lower frequencies below 100 Hz (Fig. 1c-2). This indicates that the coherence time of atmospheric turbulence lies in the millisecond range, as usually the case in normal air conditions[25]. The propagation delay of light traveling over our 18 km free-space link is 62.5 µs, much faster than the atmospheric coherence time, thereby permitting turbulent eddies to be assumed frozen in space[26]. In other words, even in the presence of scintillation, temporal atmospheric phase noise can be predicted using the weak-turbulence model[16]. Transmission break occurs when the received power drops below a threshold limit of 0.5 µW that is set by the minimum input power required for subsequent EDFA-based power amplification. The return path from Site B to Site A needed for phase noise compensation is constructed so as to maintain a high level of reciprocity with the one-way path from Site A to Site B, i.e., by employing the identical bistatic telescope pair together with the same single-mode fiber delivery and signal power amplification components.

For proof-of-concept experiments, four optical carriers are prepared with each being distinguished as $\lambda_{t,\#1}$, $\lambda_c$, $\lambda_{t,\#2}$ and $\lambda_{t,\#3}$ in increasing order of wavelength. Each carrier is sent out with 250 mW optical power. Among the four carriers, the second carrier $\lambda_c$ is selected as the feedback compensation channel of optical phase noise, which is sent back from Site B to Site A. The returned $\lambda_c$ is then beat with its original frequency to detect the Doppler shift accumulated by atmospheric phase noise during round-trip transmission. Instantaneous correction is then made by phase-locked loop (PLL) control using an acousto-optic modulator (AOM) operating at a 40 MHz nominal driving frequency. The PLL control cancels the Doppler shift on not just $\lambda_c$ but also other carriers $\lambda_{t,\#1}$, $\lambda_{t,\#2}$ and $\lambda_{t,\#3}$ by same amount. This collective scheme of undiscriminating compensation is based on the reasoning that the λ-dependent variation of the refractive index of air in the selected infrared range is not significant[27], being merely of the order of $10^{-8}$ (Methods) and smaller by three orders of magnitude than that of an equivalent fiber link (Fig. 1b). In addition, the atmospheric optical phase noise predicted by the Kolmogorov spectrum[16,26] differs within 4.5% in terms of the power spectral density (PSD) over the entire infrared range covering all the optical carriers. Conclusively, the non-common-mode noise becomes insignificant when the Doppler shift is suppressed by a signal-to-noise ratio of as large as -80 dBc/Hz from all the carriers as will be described in Fig. 3.

Figure 2 shows how the optical carriers are generated from an Er-doped fiber oscillator (C-fiber, Menlo Systems GmbH). The carriers are extracted from the source comb through a composite spectral filter and amplified by diode-based injection locking[28,29] without linewidth degradation (Methods). The source comb consists of ~ $10^5$ frequency modes evenly distributed with a 100 MHz mode spacing over a 75 nm spectral bandwidth around a 1550 nm center wavelength. The $n$-th comb mode corresponds to an optical frequency of $n \times f_r + f_o$ with $f_r$ being the repetition rate and $f_o$ the offset frequency. For comb stabilization, $f_r$ is regulated by elongating the fiber oscillator length using a piezoelectric (PZT) actuator, while $f_o$ is fixed by adjusting the pumping power after being detected using an $f$-$2f$ interferometer[29]. Then, the whole source comb is positioned in the frequency domain using an acousto-optic modulator (AOM) by tuning $\lambda_{t,\#3}$ to a high-finesse cavity using the Pound-Drever-Hall (PDH) technique[30] (Methods). The four carriers are consequently stabilized with a narrow linewidth of 1.5 Hz at 1 s averaging with their own absolute frequencies being accurately identified within the source comb (Fig. 2b). The phase noise level of the source comb is measured to be -40 dBc/Hz at 10 Hz offset after stabilization (Fig. 2c) with a slight bump at 100 kHz offset due to the PDH control bandwidth. The fractional frequency stability of the source comb is calculated from the measured phase noise to be $3.79 \times 10^{-15}$ at 0.1 s averaging time (Fig. 2d).

Figure 3 shows our experimental results of optical frequency transfer. The driving frequency of the AOM employed for phase noise compensation (Fig. 1) gives a 40 MHz shift commonly to all the optical carriers in transmission from their comb-rooted frequencies. Thus, the optical carriers arriving at Site B is evaluated by producing RF heterodyne beats at the given offset at 40 MHz with their original signals (Fig. 3a). When no phase noise compensation is implemented, no sharp RF heterodyne beats are observed (Fig. 3a). This implies that the spectral power concentrated on a 1.5 Hz narrow linewidth of each carrier is severely dispersed by atmospheric phase noise. On the other hand, when compensation control is turned on, the optical carriers regain their original linewidths as clearly observed in their RF beat spectra (Fig. 3a). Specifically, the compensation channel $\lambda_c$ recovers to a signal-to-noise ratio (SNR) of 32.1 dB, while other carriers $\lambda_{t,\#1}$ and $\lambda_{t,\#2}$ regain their SNRs to 30.3 dB at 1530.3 nm and 30.9 dB at 1564.4 nm, respectively. The restored SNRs indicate that almost 90 % of the spectral power of each carrier is brought back on its center frequency. The RF beat spectra also show that the transmitted carriers experience no notable shifts in their absolute frequency positions, confirming that no significant steady-state error arises from the PLL-based scheme of phase noise compensation described in Fig. 1.

The phase noise spectrum of the feedback channel $\lambda_c$ measured upon arrival at Site A without compensation control exhibits a strong level of atmospheric optical noise particularly at lower offset frequencies up to several hundreds of Hz (Fig. 3b). The measured power spectral density (PSD) of atmospheric noise scales to the -8/3th power of the offset frequency, showing a good agreement with the Kolmogorov spectrum of weak turbulence[16] (Methods). For higher offset frequencies from about 1 kHz, the fiber-induced white frequency noise that descends to the -2nd power of the offset frequency begins to surpass atmospheric noise. With compensation control turned on, the atmospheric noise power measured on $\lambda_c$ at Site A drops to a level below -80 dBc/Hz, while the one-way phase fluctuation of $\lambda_c$ measured at

Site B shows a slightly higher level of above -80 dBc/Hz. This relatively less noise suppression on the one-way site is unavoidable when the Doppler shift is corrected at the feedback site[9]. Nonetheless, other carriers $\lambda_{t,\#1}$ and $\lambda_{t,\#2}$ show a comparable level of one-way phase fluctuation, similar to that of $\lambda_c$, without notable difference between them.

The round-trip feedback loop - from Site A to Site B and back to Site A – causes a control time delay of 0.12 ms, which leads to a servo spike at 4.0 kHz in the phase noise spectrum (Fig. 3b). The PLL control bandwidth $f_c$ of noise compensation is set much higher at 30 kHz, but the phase noise about the servo spike is not well suppressed by the proportional-integral (PI) control used in this compensation. This causes the RF beat spectra of the transmitted carriers (Fig. 3a) to exhibit side peaks at ±4.0 kHz offset frequencies symmetrically about their center frequencies. More specifically, for one-way transmission from Site A to Site B, the integrated phase jitter from 10 Hz to the servo spike is calculated to be 0.061 rad for $\lambda_c$, being cut from 0.608 rad before noise compensation. For $\lambda_{t,\#1}$ and $\lambda_{t,\#2}$, the integrated phase jitter turns out to be 0.069 and 0.072 rad, respectively, showing no clear difference between them. The phase jitter difference between $\lambda_c$ and $\lambda_{t,\#2}$ is 0.011 rad, i.e., 18% of that of $\lambda_c$. When the optical frequency stability is evaluated in terms of the Allan deviation (Fig. 3c), it is $8.68\times10^{-15}$ at 0.1 s averaging for all the channels before compensation. After compensation, the stability is enhanced by one order of magnitude over to $3.54\times10^{-16}$ at the compensation channel $\lambda_c$, $3.99\times10^{-16}$ at $\lambda_{t,\#1}$ and $4.53\times10^{-16}$ at $\lambda_{t,\#}$. This result implies that all the carriers after phase noise compensation become stable enough to carry the state-of-the-art optical clock signals of which the frequency stability is of the order of $10^{-15}$ at 0.1 s currently[1,2].

The comb-rooted optical frequencies created in this study for optical transfer are also able to deliver microwave signals by taking advantage of the inter-comb-mode stability. In other words, heterodyning two separate comb-rooted optical carriers through a fast photodetector permits breeding well-defined higher RF frequencies up to ~100 GHz and even higher THz-waves by incorporating an appropriate photomixer[31]. In fact microwaves may also be produced from the source comb simply by electrical filtering of the RF harmonics of the pulse repetition rate[32], but the photonic way of synthesizing a microwave using two continuous optical waves as attempted here offers superior side-mode-suppression ratio (SMSR), wider frequency tunability, and higher signal-to-noise ratio (SNR). More importantly, when the undisturbed transfer of microwaves over a long-distance is necessary, synthesizing them at the receiving site by combining a pair of comb-rooted optical carriers upon transmission offers an exclusive benefit of cancelling out the atmospheric phase disturbance that equally affects both the carriers.

Figure 4 shows our experimental result of microwaves delivery in the frequency range of 0.1 to 50 GHz by varying the spectral gap between $\lambda_{t,\#2}$ and $\lambda_{t,\#3}$. Here, the phase noise compensation used for optical frequency transfer is not necessary, thus the feedback channel $\lambda_c$ is deactivated while only $\lambda_{t,\#1}$ is locked to the high-finesse cavity for stabilization of the source comb. Upon synthesis at Site B by heterodyning $\lambda_{t,\#2}$ and $\lambda_{t,\#3}$, the delivered microwave signal is filtered out through a high speed photodetector of a 100 GHz bandwidth and subsequently compared with its original microwave signal supplied from Site A. This self-heterodyning microwave test reveals that the SNR is measured as high as 40 dB while the SMSR reaches 30 dB without notable difference among four microwaves delivered (Fig. 4a). The phase noise spectrum (Fig. 4b) observed for a 10 GHz signal shows a noise level of -105 dBc/Hz at 1 Hz offset and -145 dBc/Hz at 1 kHz offset, which is slightly higher in overall than the back-to-back level measured right at the outlet of the source comb. The deviation is reckoned attributable to the fiber-induced white frequency noise arising from a total of 500 m fiber link existing in the round-trip transfer. The reference level of microwaves noise (Fig. 4b) that would be encountered by the atmospheric disturbance is estimated by downscaling the optical phase noise level (Fig. 3b) from its nominal optical frequency of 200 THz to the synthesized RF frequency of 10 GHz. The reference level is worked out higher than the measured phase noise spectrum particularly by two orders of magnitude from 5 Hz to 200 Hz offset frequencies, but it is well suppressed in the transmitted microwave even without active noise compensation control. Conclusively, the integrated phase jitter is evaluated to be 134.41 μrad for the 10 GHz microwave and 88.79 μrad for the back-to-back signal (Fig. 4b).

As an add-on demonstration, the active noise compensation system explained so far is applied to the task of coherence optical communications as illustrated in Fig. 5a. Specifically, phase-sensitive binary phase-shift keying (BPSK) transmission is implemented at 1 Gbps individually on each carrier while $\lambda_{t,\#1}$ and $\lambda_{t,\#2}$ are transmitted independently. Note that the optical link is shortened to a 1.4 km free-space path to lessen the strong level of scintillation present in the original 18 km optical link. The compensation channel of $\lambda_c$ is activated and $\lambda_{t,\#3}$ is locked to the high-finesse cavity for stabilization of the source comb. At the receiving site, the transmitted carriers are demodulated by self-homodyne detection and then digitized through a high-speed digital oscilloscope at a 10 GS/s sampling rate. Eye diagrams and constellation diagrams obtained in Fig. 5b reveal that atmospheric phase noise is well suppressed for both $\lambda_{t,\#1}$ and $\lambda_{t,\#2}$ despite their wide spectral separation of 4.2 THz. The error vector magnitude (EVM) is evaluated to be 19.4 % for $\lambda_{t,\#1}$ and 22.3 % for $\lambda_{t,\#2}$ with noise compensation on. The bit error rate (BERs) is calculated to be of the order of $10^{-10}$ or less from the measured EVM values[33]. It is anticipated that the BER would fall below $10^{-15}$ if the well-established post-processing of forward error correction (FEC) is incorporated with a 7 % overhead[34]. This result validates that our comb-rooted scheme of optical frequency transfer is capable of implementing coherent communications through the open air even without increasing hardware complexity to employ optical phase-locked loop (OPLL) or digital signal processing (DSP) of phase estimation[35,36]. The data transmission rate is expected to reach 6.4 Tbps using a single frequency comb by implementing 16-quadrature amplitude modulation (QAM) along with a 25-GHz mode spacing as well as a 10 GHz baud rate.

To conclude, our free-space transfer of optical frequencies over an 18 km outdoor link has been made successfully with suppression of atmospheric phase noise to -80 dBc/Hz. The concurrent transfer of multiple comb-rooted optical carriers has also permitted delivering microwave signals by pairing two separate carriers with inter-comb-mode phase coherence, with phase noise of -145 dBc/Hz at 1 kHz offset for a 10 GHz microwave signal. Further, coherent optical communication has been demonstrated with the potential of multi-Tbps data transmission by multiplexing comb-rooted carriers with a 25 GHz mode spacing in free space. The proposed free-space transfer of comb-rooted optical and microwave frequencies is expected to facilitate many photonic applications such as atomic clock dissemination, fundamental constants measurements, long-baseline interferometers, and massive coherent optical communications.

**Methods**

**Fine tracking mechanism.** The Cassegrain-type telescopes installed at Site A and Site B of Fig. 1 are pre-aligned with respect to their counterpart refractive transmitters using a guide beam of He-Ne laser so as to receive at least 1 mW optical power at the focal point of the telescope secondary mirror. The received light is then focused on a single-mode delivery fiber by control of a fine tracking mechanism that is comprised of a 2-D galvano mirror actuator and a quadrant position sensitive detector (PSD). Unless scintillation is extremely strong, the delivery fiber receives as much as 30 μW while the minimum threshold power required for the subsequent EDFA amplification is 0.5 μW.

**Dispersion effect on air refractive index.** The refractive index $n$ of air is larger than unity and its well-known Edlen or Ciddor empirical equation[27] is given in the form of $n - 1 \approx K(\lambda) \times D(T, P, P_{CO2}, P_w)$. The term $K$ depends on the light wavelength $\lambda$, while the term $D$ is a function of the varying environmental parameters of temperature ($T$), pressure ($P$), $CO_2$ concentration ($P_{CO2}$) and water vapor pressure ($P_w$). The $\lambda$-dependency of the dispersion term $K$ is found not large in the infrared light range, with its variation within the whole optical carrier range from $\lambda_{t,\#1}$ (1530.3 nm) to $\lambda_{t,\#3}$ (1564.4 nm) being of the order of $10^{-8}$. The atmospheric disturbance is mainly affected by $D$, which may be assumed to equally apply to $\lambda_{t,\#1}$, $\lambda_c$, $\lambda_{t,\#2}$ and $\lambda_{t,\#3}$ without significant $\lambda$-dependent non-common-mode noise. In comparison, the fiber delivery line yields relatively severe dispersion, by three orders of magnitude larger than air as depicted in Fig. 1b. Thus, the fiber delivery lines and connected optical components are kept in a temperature-controlled environment to minimize $\lambda$-dependent non-common-mode noise.

**Comb-rooted optical carrier generation.** The source frequency comb is produced from an Er-doped fiber femtosecond laser (C-fiber, Menlosystems GmbH), having a 100 MHz mode spacing over a 75 nm spectral bandwidth around a 1550 nm center wavelength. The total comb power is 20 mW with a single mode power of 500 nW. For each carrier, a band-pass filter unit is devised by combining a fiber Fabry-Perot (FFP) filter with a fiber Bragg grating (FBG). Diode-based stimulated emission by injection locking is used to amplify the extracted carrier frequency mode through a distributed feedback (DFB) laser diode. This all-fiber-based injection locking provides an amplification factor of 40 – 50 dB for each optical carrier without frequency shifting or linewidth broadening. Detailed information on this comb-based optical frequency generation is available in the authors' recent publication of Ref. 28.

**Pound-Drever-Hall control.** The high-finesse cavity acting as the optical reference provides resonance peaks of 7.5 kHz bandwidth (FWHM), corresponding to a finesse value of 400,000, with a 3.0 GHz free spectral range (FSR). The PDH control signal is obtained by phase modulation using an electro-optic modulator (EOM), of which the amplitude becomes proportional to the frequency deviation of $\lambda_{t,\#3}$ from the nearest resonance peak of the high-finesse cavity. Subsequently, the PDH control signal is nullified by laterally shifting the optical spectrum of the source comb using an acousto-optic modulator (AOM) and at the same time by controlling the pulse repetition rate using a PZT actuator installed to vary the cavity length of the source comb. Simultaneously, the offset frequency $f_o$ is adjusted to nullify the AOM-induced lateral shift of the whole comb, which is made by adjusting the pumping power after being detected using an *f*-2*f* interferometer. Another frequency comb is stabilized to the same high-finesse cavity as the reference comb, from which the beat linewidth of each optical carrier of Fig. 2b is confirmed to be 1.5 Hz as detailed in Ref. 28.

**Atmospheric turbulence disturbance.** The Kolmogorov spectrum[16,26] yields the power spectral density (PSD) of atmospheric optical phase noise in terms of the noise frequency $f$ as $S_\Phi(f)=0.016k^2C_n^2LV^{5/3}f^{-8/3}$ [rad$^2$/Hz]; with $k$ being the wavenumber of the optical carrier under transmission, $C_n^2$ the air refractive-index structure parameter, $L$ the transmission distance and $V$ the wind speed perpendicular to the transmission direction. The $\lambda$-dependence dispersion effect on $S_\Phi(f)$ is just proportional to the term $k^2$, but its variation within the optical carrier range (35 nm) is restricted to 4.5%. Our experimental conditions predict $C_n^2$ to be $2\times10^{-14}$ m$^{-2/3}$ for $L$=18 km and $V$=1 m/s.

**Acknowledgments**

This work was supported by the National Research Foundation of the Republic of Korea (NRF-2012R1A3A1050386). Y.-J. Kim acknowledges support from the Singapore National Research Foundation (NRF-NRFF2015-02).

**Author contributions**

The project was planned and overseen by S.-W.K. in collaborations with Y.-J.K. Experiments were performed by H.J.K. and J.-W.Y. with supports of B.J.C., H.J., and B.S.K. All authors contributed to the manuscript preparation.

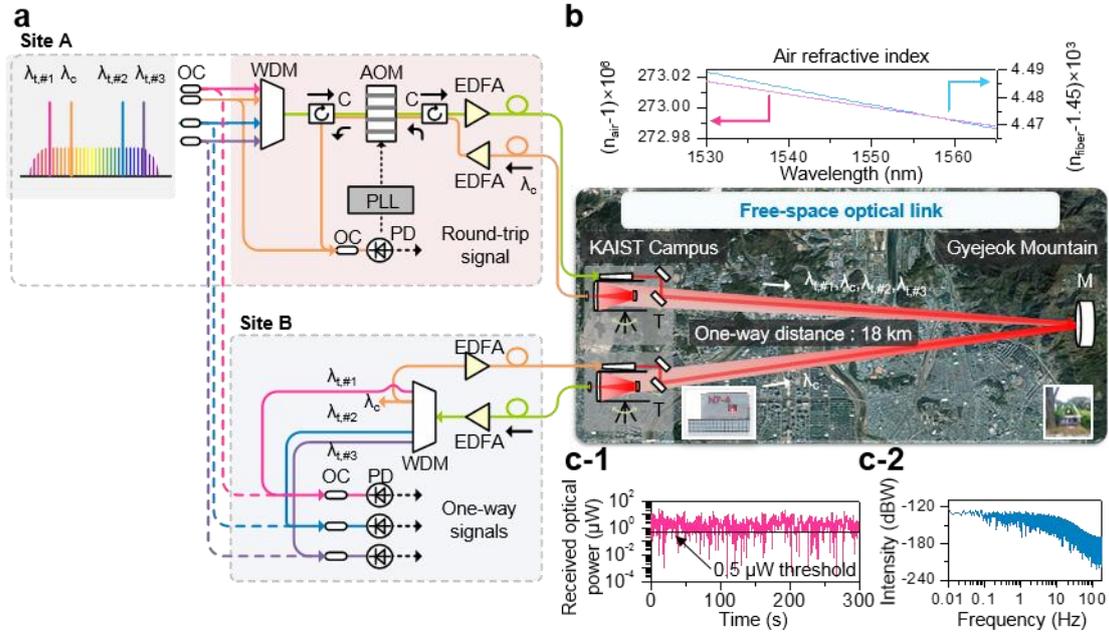

**Figure 1 | Phase-coherent transfer of comb-rooted optical frequencies over an 18-km free space link. a**, Multiple optical carriers are extracted from a frequency comb in stabilization to a high-finesse cavity and transmitted from Site A to Site B. The second carrier $\lambda_c$ is selected as the feedback compensation channel of atmospheric optical phase noise, while other carriers $\lambda_{t,\#1}$, $\lambda_{t,\#2}$ and $\lambda_{t,\#3}$ are dedicated channels for optical frequency transfer. **b**, Refractive indices of air and fiber in the infrared range are compared in the inset. **c-1**, Optical power scintillation at Site B observed at a sampling rate of 3 ms over an interval of 300 s. **c-2**, Scintillation frequency spectrum. Abbreviations; AOM: acousto-optic modulator, C: optical circulator, EDFA: erbium-doped fiber amplifier, PLL: phase-locked loop, T: telescope, M: flat mirror, OC: optical coupler, PD: photodetector, WDM: wavelength division multiplexer.

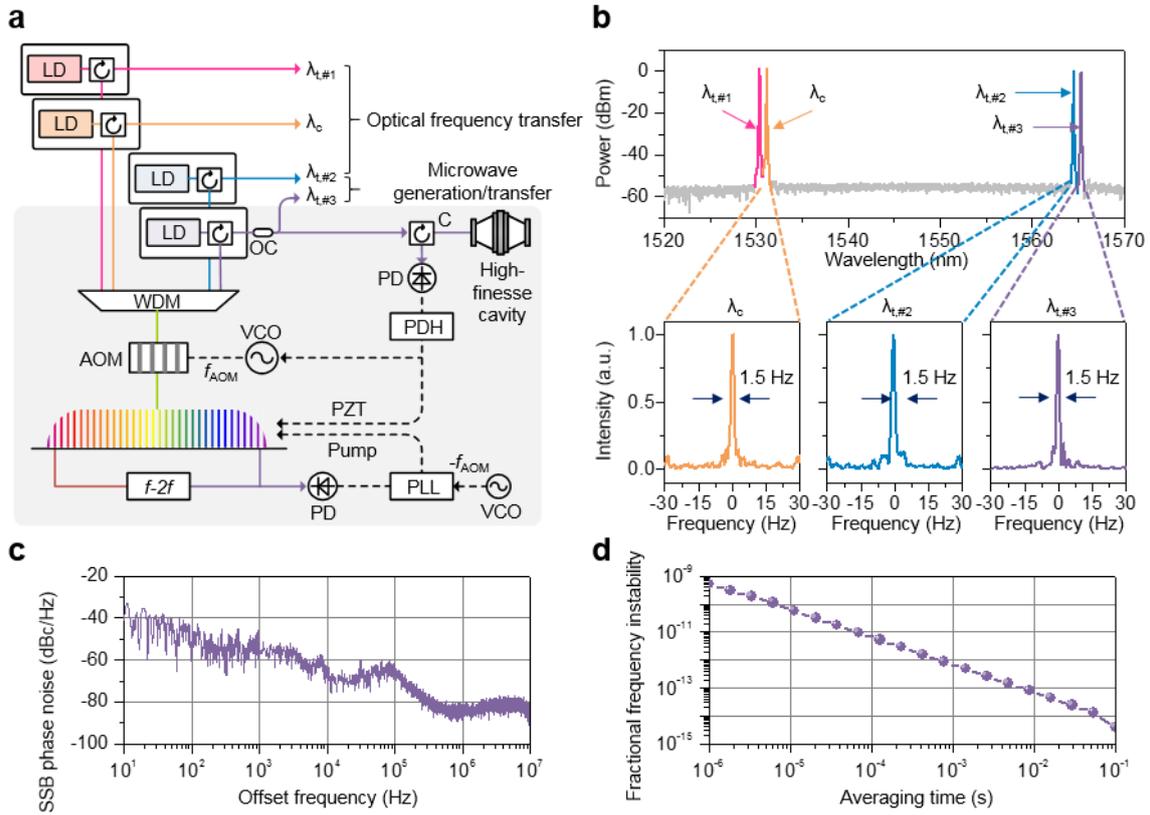

**Figure 2 | Comb-rooted optical frequency synthesis. a**, The source comb is stabilized by controlling the repetition rate and the offset frequency. Multiple comb modes are selected out as optical carriers and amplified by injection locking to laser diodes (LD), while one comb mode is phase-locked directly to a high-finesse cavity. PD: photodetector, C: optical circulator, OC: optical coupler, VCO: voltage-controlled oscillator, AOM: acousto-optic modulator, PZT: piezo-electric transducer. WDM: wavelength division multiplexer. **b**, Optical and RF beat spectra of the optical carriers selected from 1530.3 nm to 1564.4 nm, all showing a 1.5 Hz instrument-limited narrow linewidth. **c**, Single side band (SSB) phase noise spectrum of the source comb after stabilization to the high-finesse cavity. **d,** The source comb's Allan deviation that is converted from the SSB phase noise spectrum.

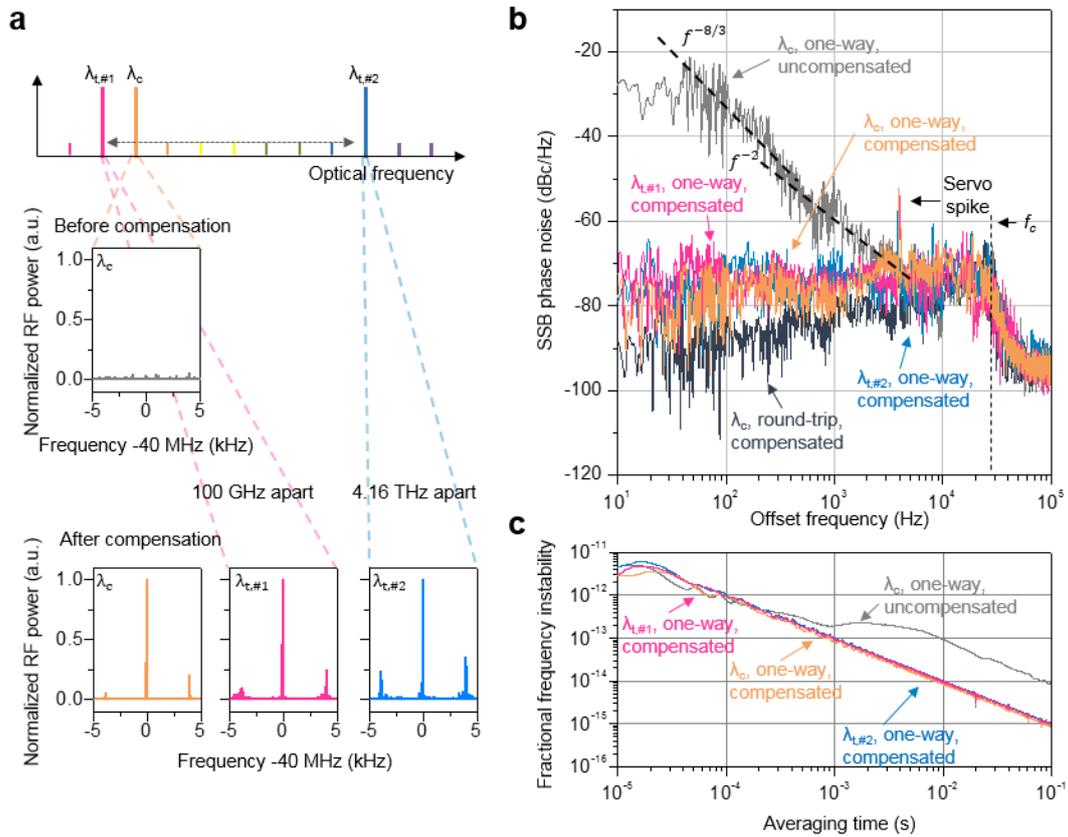

**Figure 3 | Coherent transfer of optical frequencies over an 18-km free-space link. a**, RF beats of comb-rooted carriers with original carrier signals monitored before and after phase noise compensation. **b**, Single side band (SSB) phase noise spectra of one-way (form Site A to Site B) and round-trip (Site A – Site B – Site A) optical carriers. $f_c$ denotes the control bandwidth of phase noise compensation. **c**, Allan deviation of one-way optical carriers after phase noise compensation.

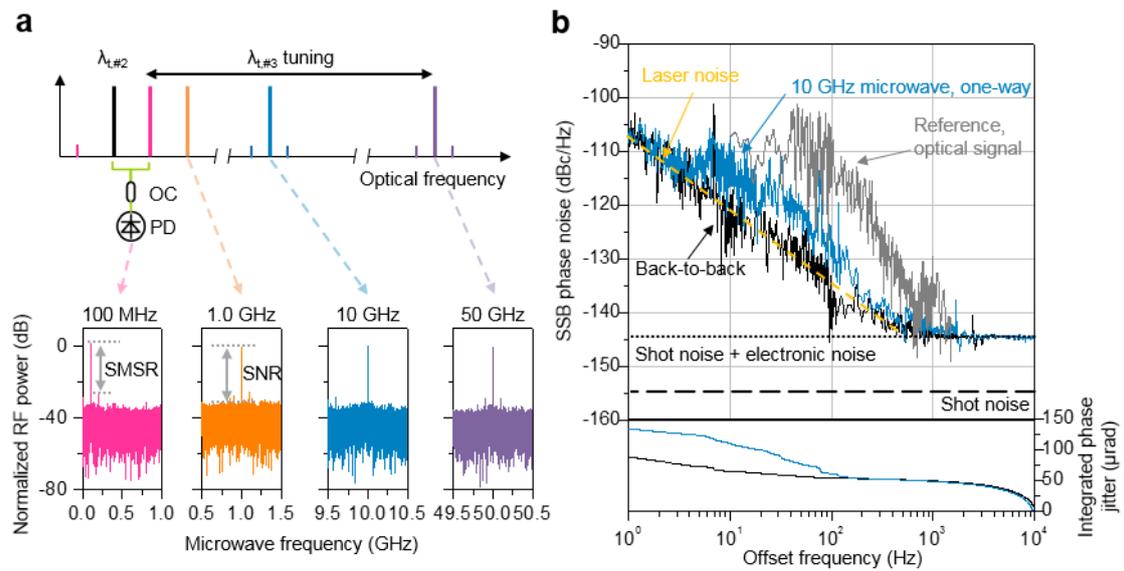

**Figure 4 | Microwave signal transfer. a**, Microwave synthesis by heterodyning two optical carriers $\lambda_{t\#2}$ and $\lambda_{t\#3}$. Abbreviations are; OC: optical coupler, PD; photodetector, SMSR: side-mode-suppression ratio, SNR: signal-to-noise ratio. **b**, Phase noise spectra of a 10 GHz microwave measured at Site A (black) in back-to-back configuration and at Site B (blue) after transmission over an 18 km free-space optical link. The reference optical signal (grey) is plotted by downscaling the uncompensated optical phase noise shown in Fig. 3b.

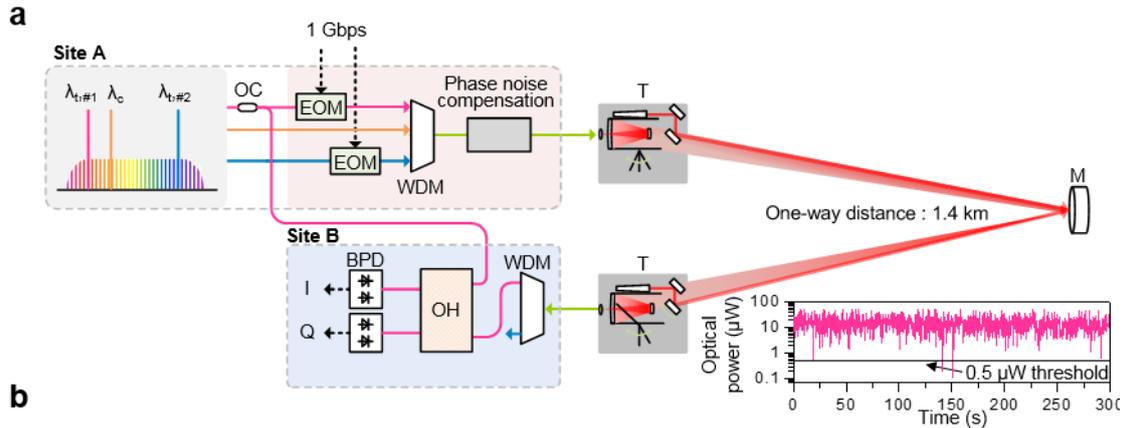

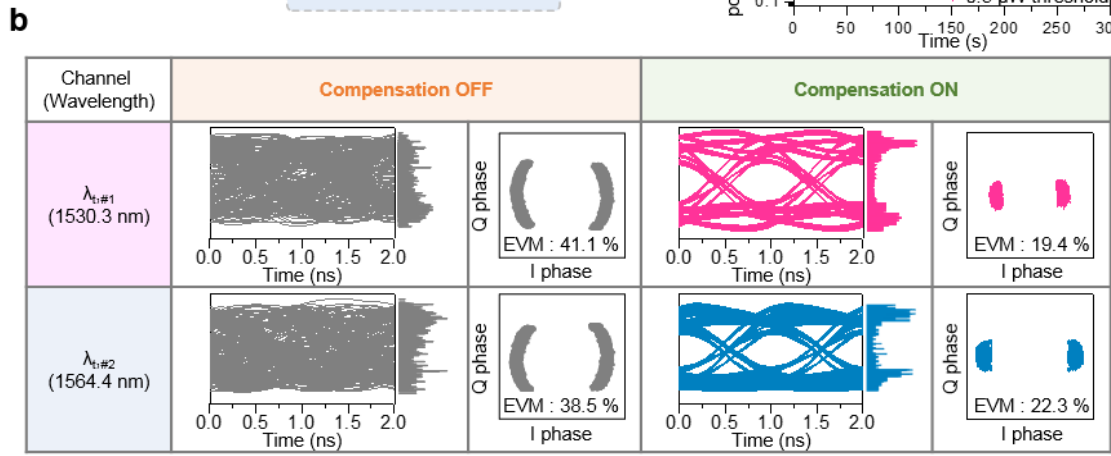

**Figure 5 | Coherent optical communication test over a shortened link of a 1.4 km distance. a**, Binary phase-modulation is made at a 1 Gbps transmission rate using an electro-optic modulator (EOM). At the receiving site, incoming carrier signals are de-multiplexed through a wavelength division multiplexer (WDM), demodulated using an optical hybrid (OH) device, and finally converted by balanced photo-detectors (BPDs) to the in-phase (I) and quadrature-phase (Q) signals. (Inset) The measured optical power is for $\lambda_c$ transmitted with 100 mW power. **b**, Performance evaluations in terms of eye diagrams, constellation diagrams, and error vector magnitudes (EVMs) before and after phase noise compensation.